\documentclass{article}
\usepackage{amsmath,graphicx}
\usepackage{amssymb,epsfig}
\usepackage{tikz,booktabs,float}
\usetikzlibrary{arrows}
\usepackage[utf8]{inputenc}
\usepackage{pgfplots}
\pgfplotsset{width=10cm,compat=1.9}
\usepackage[preprint]{spconf}

\title{MULTILINGUAL ADAPTATION OF RNN BASED ASR SYSTEMS}
\name{Markus Müller, Sebastian Stüker, Alex Waibel \thanks{This work was realized in the framework of the ANR-DFG project BULB (ANR-14-CE35-002).}}
\address{Institute for Anthropomatics and Robotics  \\
Karlsruhe Institute of Technology, Karlsruhe, Germany \\
{\small \tt \{m.mueller, sebastian.stueker, alexander.waibel\}@kit.edu}
}
\begin{document}
\copyrightnotice{\copyright\ IEEE 2018}
\toappear{To appear in {\it Proc. ICASSP 2018, April 15-20, 2018, Calgary, Canada}}
\maketitle
\begin{abstract}
In this work, we focus on multilingual systems based on recurrent neural networks (RNNs), trained using the Connectionist Temporal Classification (CTC) loss function.
Using a multilingual set of acoustic units poses difficulties.
To address this issue, we proposed Language Feature Vectors (LFVs) to train language adaptive multilingual systems.
Language adaptation, in contrast to speaker adaptation, needs to be applied not only on the feature level, but also to deeper layers of the network.
In this work, we therefore extended our previous approach by introducing a novel technique which we call ``modulation''.
Based on this method, we modulated the hidden layers of RNNs using LFVs.
We evaluated this approach in both full and low resource conditions, as well as for grapheme and phone based systems.
Lower error rates throughout the different conditions could be achieved by the use of the modulation.
\end{abstract}
\section{Introduction}
\label{sec:intro}
Training multilingual speech recognition systems requires special methods.
In low-resource conditions, training systems on data from multiple languages improves the performance.
In a resource rich environment, using data from multiple languages often does not improve the performance, but might event affect it negatively.
In both cases, adaptation techniques are required to improve the recognition accuracy and neural networks adapted to language characteristics have proven to perform better.
This is similar to speaker adaptation, where adapted networks outperform unadapted ones.

However, language adaptation is more challenging than speaker adaptation:
Collecting training data from several hundred speakers is possible.
This amount of speakers enables networks to generalize upon speaker properties.
For language adaptation, there are an order of magnitude less languages available than there are speakers.
This renders generalization across languages more difficult.
Another factor is the task itself.
When trained on data from multiple speakers of the same language, the same targets, e.g., phone states, are used.
Different languages feature different, but sometimes overlapping, sets of targets.
Although speech recognition for different languages are different tasks, they are related since all languages are being spoken by humans.
This limits the sound inventory to the sounds that can be produced by the human vocal tract.
Also, languages (from the same language family) potentially share sound inventories, as well as the set of targets the network is trained on.

Applying language adaptation techniques should therefore enable the networks to generalize better.
Encoding language properties using, e.g., LFVs, like we showed in the past, allow networks to be trained language adaptive in such a way that they can exploit similarities and differences between languages.
Unlike traditional GMM/HMM or DNN/HMM based systems, RNN/CTC based setups do not require explicit modelling of context dependent states which would then need to be adapted.
Based on RNNs, these systems should be trained towards learning features based on language properties in order to be able to better perform in a multilingual scenario.

As we outlined in the related works section, several techniques for language adaptation have been proposed for traditional systems in the past.
We proposed to use LFVs as additional input features for language adaptation.
In this paper, we introduce a novel approach of integrating LFVs into recurrent network architectures based on the idea of Meta-PI networks.
The effectiveness of our approach is demonstrated in a series of experiments, showing that the method presented here can be applied to both full- and low-resource conditions.
In addition, we also omitted the pronunciation dictionary and built systems using graphemes only.
In a multilingual scenario, this is particularly challenging as the network is required to learn pronunciations from multiple languages in parallel.
To evaluate our systems, we use the token error rate (TER) as primary measure of the trained networks.
But we also incorporated a RNN based language model (LM) for decoding to the determine the word error rate (WER).

This paper is organized as follows: In the next Section \ref{sec:relwork}, we outlined related work in the field, followed by a detailed description of the method proposed in Section \ref{sec:adapt}.
We described the experimental setup in Section \ref{sec:expsetup}, followed by the results of our experiments (Section \ref{sec:results}). This paper concludes with Section \ref{sec:conclusion}, where we also outline possible future work.
\section{Related Work}
\label{sec:relwork}
\subsection{Multi- and Crosslingual Speech Recognition Systems}
\label{sec:relwork:subsec:gmm}
Prior to the emergence of neural networks, ASR systems were typically built using a GMM/HMM based approach.
Methods for training/adapting such systems cross- and multilingually were proposed to handle data sparsity \cite{schultz1997fast,stuker2009acoustic}.
The process of clustering context-independent phones into context-dependent ones can also be adapted to account for cross- and multilinguality \cite{schultz2000polyphone}.
Due to their recurrent nature, RNNs are a powerful tool to model sequential dependencies, rendering the need for context-dependent targets superfluous.
Using only context-independent targets has the advantage that no clustering is required.
\subsection{Multilingual Bottleneck Features}
In a resource constraint scenario, data from additional source languages are used to improve the performance.
DNNs are typically trained in two steps: Pre-training and fine-tuning.
It has been shown that the pre-training step is language independent \cite{SwietojanskiGR12}.
The fine-tuning can be modified in multiple ways to account for additional languages.
One approach includes the use of shared hidden layers, with language dependent output layers \cite{VeselyKGJE12}.
Combining multiple output layers into one is also possible \cite{grezl2014adaptation}.
\subsection{Neural Network Adaptation}
Feeding additional input features into a neural network is a common way for adaptation.
A popular approach for speaker adaptation is to supply i-Vectors \cite{saon2013speaker}, which encode speaker characteristics in a low-dimensional representation.
Speaker adaptive neural networks can be trained this way \cite{miao2014towards}.
Such low dimensional codes can also be extracted using neural networks, called Bottleneck Speaker Vectors (BSVs) \cite{huang2015investigation}.
In the past, we proposed similar methods to adapt DNNs to multiple languages.
We first introduced a method encoding the language identity using one-hot encoding \cite{mueller2015}.
We enhanced this method in a similar way to BSVs, by extracting Language Feature Vectors (LFVs) \cite{mueller2016}.
These vectors have shown to encode language properties instead of the language identity alone, even for languages not seen during training.
\subsection{RNN Based ASR Systems}
RNN based ASR systems are becoming increasingly popular.
One method to train them is the use of the Connectionist temporal classification (CTC) loss function \cite{graves2006connectionist}, which does not require frame-level labels.
It aligns a sequence of tokens automatically.
As in traditional systems, phones, graphemes or both combined can be used as acoustic modeling units \cite{chen2014joint}.
Given enough training data, even whole words can be used \cite{soltau2016neural}.
\section{Language Adaptation}
\label{sec:adapt}
In the past, we proposed methods for adapting multilingual neural network based ASR systems to languages using LFVs.
Language Feature Vectors are a low-dimensional representation of language properties, extracted via a neural network.
This network was trained to discriminate languages, based on log Mel and tonal features (FFV \cite{kornel:ffv} and pitch \cite{kjell:da}) typically used by ASR systems.
A similar architecture as for extraction of BNFs was used.
This architecture featured a bottleneck as second last layer.
After training, the output activations of this layer were used as LFVs.
To perform the adaptation, we appended LFVs to the acoustic features, similar to appending i-Vectors for speaker adaptation.
In the results section, we included error rates using this method as contrastive experiments, denoted as ``LFV app''.
This method has shown to reduce error rates for multilingual GMM/HMM, DNN/HMM as well as RNN/CTC based systems.
\subsection{Neural Modulation}
Appending features for speaker adaptation to acoustic features is fitting, as changes in speaker characteristics are reflected within the signal.
Multiple adaptation methods like VTLN or fMLLR which directly operate on the acoustic features were proposed.
The same holds true for i-Vector based adaptation, where speaker adaptive systems can be trained to directly shift input features based on speaker properties \cite{miao2014towards}.
But language properties are a higher order concept in contrast to speaker variations.
Some aspects are based on acoustics, e.g. having the same phone in multiple languages, where a language specific coloring can be observed to some degree.
But aspects like phonotactics or different sets of acoustic units require adaptation methods beyond the transformation of acoustic features.
Here, adding features at deeper layers potentially enables better adaptation.

One possibility is a method first introduced as part of Meta-PI \cite{hampshire1992meta} networks.
The key aspect is the use of Meta-PI connections, which allow to modulate the output of units by multiplication with a coefficient.
Applied to language adaptation, we modulated the outputs of hidden layers with LFVs.
Based on language features, the output of LSTM cells are attenuated or emphasized.
This forces the cells in the hidden layer to learn or adapt to features based on language properties.
Modulation can be considered related to dropout training \cite{hinton2012improving}, where connections are dropped on a random basis.
In the results section, we refer to this method as ``LFV mod''.

We used a network configuration as shown in Figure \ref{fig:ctcnn}.
The basic architecture is inspired by Baidu's Deepspeech 2.
It combines two TDNN/CNN layers with 4 bi-directional LSTM layers.
The output layer is a feed-forward layer which maps the output of the last LSTM layer to the targets.
We chose the number of LSTM cells in each layer to be a multiple of the dimensionality of the LFVs.
This way, we could structure the hidden layer into groups of LSTM cells containing an equal amount of units.
The output of each group is then modulated with one dimension of the LFVs.
The figure shows both configurations, ``LFV app'' and ``LFV mod'', but only one method was applied at a time.
In preliminary experiments, we determined modulating the output of the second LSTM layer to result in the best performance.
\tikzstyle{layer}=[draw=black,fill=black!30]
\tikzstyle{layerlfv}=[draw=black,fill=green!70]
\tikzstyle{layerrnn}=[draw=black,fill=blue!20]
\tikzstyle{layercnn}=[draw=black,fill=orange!30]
\tikzstyle{arrow} = [semithick,fill=red!30,line width=1.4pt, shorten >= 4.5pt]
\tikzstyle{dots}=[draw=black,fill=black]
\begin{figure}[!h]
\centering
\begin{tikzpicture}[scale=0.5]

% CNN layer
\fill[layercnn] (-6,3) coordinate(c1tl) -- (-7,3) coordinate(c1tr) -- (-7,-3) coordinate(c1br) -- (-6,-3) coordinate(c1bl) -- (-6,3);

\fill[layercnn] (-4,2.5) coordinate(c2tl) -- (-5,2.5) coordinate(c2tr) -- (-5,-2.5) coordinate(c2br) -- (-4,-2.5) coordinate(c2bl) -- (-4,2.5);

% LFV layer
\fill[layerlfv] (-4,3.5) coordinate(lfvtl) -- (-5,3.5) coordinate(lfvtr) -- (-5,2.5) coordinate(lfvbr) -- (-4,2.5) coordinate(lfvbl) -- (-4,3.5);

% LSTM layer
\fill[layerrnn] (-2.5,4) coordinate(l1tl) -- (-1.5,4) coordinate(l1tr) -- (-1.5,-4) coordinate(l1br) -- (-2.5,-4) coordinate(l1bl) -- (-2.5,4);

\fill[layerrnn] (-0.5,4) coordinate(l2tl) -- (0.5,4) coordinate(l2tr) -- (0.5,-4) coordinate(l2br) -- (-0.5,-4) coordinate(l2bl) -- (-0.5,4);

\fill[layerlfv] (0.5,4) coordinate(l2atl) -- (1.0,4) coordinate(l2atr) -- (1.0,-4) coordinate(l2abr) -- (0.5,-4) coordinate(l2abl) -- (0.5,4);

\fill[layerrnn] (1.5,4) coordinate(l3tl) -- (2.5,4) coordinate(l3tr) -- (2.5,-4) coordinate(l3br) -- (1.5,-4) coordinate(l3bl) -- (1.5,4);

\fill[layerrnn] (3.5,4) coordinate(l4tl) -- (4.5,4) coordinate(l4tr) -- (4.5,-4) coordinate(l4br) -- (3.5,-4) coordinate(l4bl) -- (3.5,4);

% Output layer
\fill[layer] (6,2) coordinate(outtl) -- (7,2) coordinate(outtr) -- (7,-2) coordinate(outbr) -- (6,-2) coordinate(outbl) -- (6,2);

% Descriptions
\node[align=center,font=\small,rotate=0] at (-5.5,-5) {2D TDNN / CNN\\Layers};
\node[align=center,font=\small,rotate=0] at (1,-5) {Bi-directional LSTM Layers};
\node[align=center,font=\small,rotate=0] at (6.5,-5) {Output\\Layer};
\node[align=center,font=\small,rotate=0] at (-4.5,5.0) {LFV app};
\node[align=center,font=\small,rotate=0] at (0.75,5.0) {LFV mod};

% Links between networks

% c1 - c2
\draw (c1bl) -- (c2br);
\draw (c1tl) -- (c2tr);

% c2 - l1
\draw (c2bl) -- (l1tl);

% lfv - l1
\draw (lfvtl) -- (l1bl);

% l1 - l2
\draw (l1br) -- (l2tl);
\draw (l1tr) -- (l2bl);

% l2 - l3
\draw (l2abr) -- (l3tl);
\draw (l2atr) -- (l3bl);

% l3 - l4
\draw (l3br) -- (l4tl);
\draw (l3tr) -- (l4bl);

% l4 - out
\draw (l4br) -- (outtl);
\draw (l4tr) -- (outbl);

\end{tikzpicture}
\caption{Network architecture showing ``LFV app'', as well as the proposed adaptation method ``LFV mod''.}
\label{fig:ctcnn}
\end{figure}
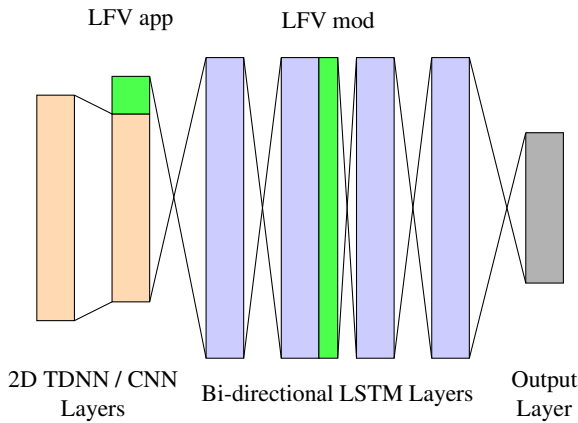
\section{Experimental Setup}
\label{sec:expsetup}
We based our experiments on the Euronews corpus \cite{gretter2014euronews}, which contains data from 10 languages.
For each language, 70h of TV broadcast news recordings are available.
For our experiments, we used a combination of 4 languages (English, French, German, Turkish), based on the availability of pronunciation dictionaries.
We filtered utterances based on length, omitting very short ones ($<$ 1s), and also removed ones having a transcript of more than 639 characters\footnote{Internal limitation within the implementation of CUDA/warp-ctc, see: https://github.com/baidu-research/warp-ctc, accessed 2018-02-12}.
Noises were only annotated in a very basic way with a single noise marker covering all different noise types, ranging from music, background and human noises.
We therefore omitted utterances marked as noise.
After applying all filtering, approx. 50h of data remained per language and was split into 45h of training and 5h of test data.
For training, we created an additional subset containing only 8h out of the 45h training set to evaluate our approach in a low-resource condition.
\subsection{Acoustic Units}
As acoustic units, we used both phones and graphemes.
The pronunciation dictionaries were created using MaryTTS \cite{schroder2003german}.
For merging the monolingual dictionaries, we mapped the phone-symbols to a multilingual phone set using the definition of articulatory features in MaryTTS' language description files.
In addition to systems based on phones, we also trained networks using graphemes as acoustic units.
To indicate word boundaries, an additional token was used.
\subsection{RNN/CTC Network Training}
Multilingual Bottleneck Features (ML-BNFs) were used as input features.
The ML-BNFs network was trained using data from 5 languages (French, German, Italian, Russian, Turkish).
Input features to the network were log Mel and tonal features (FFV \cite{kornel:ffv} and pitch \cite{kjell:da}), extracted using a 32ms window with a 10ms frame-shift.
The RNN network was trained using stochastic gradient descent (SGD) and Nesterov momentum \cite{sutskever2013importance} with a factor of $0.9$.
Mini-batch updates with a batch size of 15 were applied together with batch normalization.
The utterances were sorted ascending by length to stabilize the training, as shorter utterances are easier to align.
\subsection{Grapheme Based RNN LM}
We used a RNN based LM, trained on graphemes as described in \cite{zenkel2017comparison}.
It featured 1 hidden layer with 1024 LSTM cells.
The model was trained on only a very limited set of sentences, consisting of the training utterances of the acoustic model only.
As language models are typically trained on several millions of sentences, this is not much training data.
But the model should provide an indication whether the improvements observed as TER also result in a better word level speech recognition system.
\subsection{Evaluation}
We evaluated our proposed method varying two conditions: The availability of a pronunciation dictionary and the amount of data.
An ASR system without language adaptation is used as baseline.
First, we used the token error rate (TER) as primary measure to determine the performance without the use of external (language) models.
For decoding, we use the same procedure as in \cite{graves2006connectionist} and greedily search for the best path.
In addition to the TER, we also determined the word error rate (WER) using an RNN LM.
\section{Results}
\label{sec:results}
\subsection{Grapheme Based Systems}
First, we evaluated the use of graphemes as acoustic modeling units.
We started using a network configuration with the RNN part having 420 LSTM cells per layer, trained using only 8h of data per language (see Table \ref{tab:gra420low}).
Adding LFVs after the TDNN / CNN layers (``LFV app'') does lower the TER, but applying the method presented here (``LFV mod'') lowers the TER even more.
\begin{table}[h!]
\centering
\begin{tabular}{l|c|c|c|c}
\toprule
\textbf{Condition} & \textbf{DE} & \textbf{EN} & \textbf{FR} & \textbf{TR} \\
\midrule
ML Baseline    & 30.8 & 38.0 & 29.4 & 30.9 \\
LFV app   & 22.9 & 33.3 & 27.3 & 21.3 \\
LFV mod   & 20.7 & 32.7 & 25.4 & 19.6 \\
\bottomrule
\end{tabular}
\caption{TER of grapheme based system trained on 8h per language, 420 LSTM cells per layer}
\label{tab:gra420low}
\end{table}
Similar gains can be observed using the full training set (Table \ref{tab:gra420high}).
The use of more data lowered the TER, whereas the relative improvements were in the same order of magnitude.
\begin{table}[h!]
\centering
\begin{tabular}{l|c|c|c|c}
\toprule
\textbf{Condition} & \textbf{DE} & \textbf{EN} & \textbf{FR} & \textbf{TR} \\
\midrule
Baseline    & 10.6 & 18.2 & 15.9 & 9.1 \\
LFV app   & 9.5 & 16.1 & 14.3 & 8.1 \\
LFV mod   & 9.1 & 15.5 & 13.6 & 8.0 \\
\bottomrule
\end{tabular}
\caption{TER of grapheme based system trained on 45h per language, 420 LSTM cells per layer}
\label{tab:gra420high}
\end{table}
Training on more data also allowed for larger networks.
In an additional experiment, we increased the number of LSTM cells per layer to 840.
As shown in Table \ref{tab:gra840high}, the TER decreases in absolute terms, but the difference between addition and modulation becomes smaller.
\begin{table}[h!]
\centering
\begin{tabular}{l|c|c|c|c}
\toprule
\textbf{Condition} & \textbf{DE} & \textbf{EN} & \textbf{FR} & \textbf{TR} \\
\midrule
Baseline    & 8.9 & 15.0 & 13.5 & 7.9 \\
LFV app   & 7.9 & 13.6 & 11.8 & 7.1 \\
LFV mod   & 7.7 & 13.3 & 11.7 & 7.1 \\
\bottomrule
\end{tabular}
\caption{TER of grapheme based system trained on 45h per language, 840 LSTM cells per layer}
\label{tab:gra840high}
\end{table}
\subsection{Phoneme Based Systems}
In the same notion as graphemes, we evaluated systems based on phonemes as acoustic modelling units.
Starting with the limited data set (Table \ref{tab:phn420low}), improvements by the modulation (``LFV mod'') over the addition (``LFV app'') can be observed.
\begin{table}[h!]
\centering
\begin{tabular}{l|c|c|c|c}
\toprule
\textbf{Condition} & \textbf{DE} & \textbf{EN} & \textbf{FR} & \textbf{TR} \\
\midrule
Baseline  & 21.7 & 27.2 & 23.9 & 21.6 \\
LFV app   & 20.9 & 26.4 & 21.3 & 19.5 \\
LFV mod   & 19.0 & 25.6 & 19.8 & 17.6 \\
\bottomrule
\end{tabular}
\caption{TER of phoneme based system trained on 8h per language, 420 LSTM cells per layer}
\label{tab:phn420low}
\end{table}
Using all available training data and increasing the number of LSTM cells per layer to 840, similar improvements could be achieved (Table \ref{tab:phn840high}).
In contrast to the grapheme based setup (Table \ref{tab:gra840high}), modulating the layers (``LFV mod'') improves the performance over the simple addition (``LFV app'').
The TERs of the grapheme based systems for German and Turkish are lower compared to their phone based counterparts.
One reason for this is the quality of the pronunciation dictionary, which was created fully automatically based on a TTS system.
\begin{table}[h!]
\centering
\begin{tabular}{l|c|c|c|c}
\toprule
\textbf{Condition}  & \textbf{DE} & \textbf{EN} & \textbf{FR} & \textbf{TR} \\
\midrule
Baseline    & 9.6 & 14.6 & 12.1 & 8.5 \\
LFV app   & 9.3 & 13.2 & 10.8 & 7.7 \\
LFV mod   & 8.6 & 12.5 & 10.2 & 7.3 \\
\bottomrule
\end{tabular}
\caption{TER of phoneme based system trained on 45h per language, 840 LSTM cells per layer}
\label{tab:phn840high}
\end{table}
\subsection{Decoding with RNN LM}
To determine the WER, we ran a greedy decoding using a char based RNN LM on the English subset of the test data.
The results shown in Table \ref{tab:gra420nnlm} indicate that the improvements of TER are also observable w. r. t. WER after decoding with a language model.
\begin{table}[h!]
\centering
\begin{tabular}{c|c|c|c}
\toprule
\textbf{Setup} & \textbf{Baseline} & \textbf{LFV app} & \textbf{LFV mod}\\
\midrule
8h-420    & 32.4\% & 30.6\% & 29.9\% \\
45h-840    & 29.2\% & 27.7\% & 27.3\% \\
\bottomrule
\end{tabular}
\caption{WERs for English grapheme based systems.}
\label{tab:gra420nnlm}
\end{table}
\section{Conclusion}
\label{sec:conclusion}
Unlike speaker adaptation, where the collection of data covering hundreds of speakers is feasible, collecting data from that many languages is next to impossible.
Optimizing the adaptation method is therefore key to maximize the performance in a multilingual scenario.
We presented an improved method for language adaptation of RNNs in a multilingual setting.
Modulating the outputs of a layer showed improvements over appending LFVs to input features.
\bibliographystyle{IEEEbib}
\bibliography{refs}

\end{document}